\documentclass{article}

\usepackage{PRIMEarxiv}
\usepackage{multicol}
\usepackage[utf8]{inputenc} 
\usepackage[T1]{fontenc}    
\usepackage{hyperref}       
\usepackage{url}            
\usepackage{booktabs}       
\usepackage{amsfonts}       
\usepackage{nicefrac}       
\usepackage{microtype}      
\usepackage{lipsum}
\usepackage{fancyhdr}       
\usepackage{graphicx}       
\usepackage{capt-of}
\usepackage{caption}
\usepackage{subcaption}
\graphicspath{{media/}} 
\usepackage{amsmath}
\usepackage{amsfonts}
\usepackage{amssymb}

\usepackage[first=0,last=9]{lcg}

\usepackage{color, colortbl}
	
\definecolor{Gray}{gray}{0.9}
\title{Modeling T1 Resting-State MRI Variants Using Convolutional Neural Networks in Diagnosis of OCD}

\author{
  Tarun Eswar \\
  Author \\
  Massachusetts Academy of Math and Science \\
  Worcester, MA \\
  \texttt{teswar@wpi.edu} \\
   \And
  Nicholas Medeiros \\
  Advisor \\
  Massachusetts Academy of Math and Science \\
  Worcester, MA \\
  \texttt{nmedeiros@wpi.edu} \\
}
\newenvironment{Figure}
  {\par\medskip\noindent\minipage{\linewidth}}
  {\endminipage\par\medskip}
\begin{document}
\maketitle
\begin{abstract}
\textnormal{Obsessive-compulsive disorder (OCD) presents itself as a highly debilitating disorder. The disorder has common associations with the prefrontal cortex and the glutamate receptor known as Metabotropic Glutamate Receptor 5 (mGluR5). This receptor has been observed to demonstrate higher levels of signaling from positron emission tomography scans measured by its distribution volume ratios in mice. Despite this evidence, studies are unable to fully verify the involvement of mGluR5 as more empirical data is needed. Computational modeling methods were used as a means of validation for previous hypotheses involving mGluR5. The inadequacies in relation to the causal factor of OCD were answered by utilizing T1 resting-state magnetic resonance imaging (TRS-MRI) scans of patients suffering from schizophrenia, major depressive disorder, and obsessive-compulsive disorder. Because comorbid cases often occur within these disorders, cross-comparative abilities become necessary to find distinctive characteristics. Two-dimensional convolutional neural networks alongside ResNet50 and MobileNet models were constructed and evaluated for efficiency. Activation heatmaps of TRS-MRI scans were outputted, allowing for transcriptomics analysis. Though, a lack of ability to predict OCD cases prevented gene expression analysis. Across all models, there was an 88.75\% validation accuracy for MDD, and 82.08\% validation accuracy for SZD under the framework of ResNet50 as well as novel computation. OCD yielded an accuracy rate of around 54.4\%. These results provided further evidence for the \textit{p-factor} theory regarding mental disorders. Future work involves the application of alternate transfer learning networks than those used in this paper to bolster accuracy rates. }
\end{abstract}

\keywords{Obsessive-compulsive disorder \and magnetic resonance imaging \and major depressive disorder \and schizophrenia}
\newpage
\begin{multicols}{2}
\section{Introduction}
\textnormal{Over recent decades, obsessive-compulsive disorder (OCD) has been ranked as one of the ten most disabling disorders \cite{murray_global_1996}. A patient suffering from OCD will often experience a variety of symptoms that fall into two main categories: obsessions and compulsions. Obsessions refer to being overly focused on a specific issue, involving overthinking in the form of impulsions. Furthermore, compulsions reflect specific actions to counteract the obsessive symptoms. These habits can include checking and mental compulsions, though the list of specific compulsions varies by case \cite{ocdf1}. The symptoms themselves seem relatively benign; however, a concern arises in relation to how much time the disorder occupies in a sufferer’s daily life. The fear associated with failing to fulfill an impulse is the major component in the factors that push a diagnosed patient to follow their obsession(s). As a result, a sufferer of OCD typically spends an hour or more each day fixated on these debilitating symptoms \cite{noauthor_nimh_nodate}. Those facing extreme cases of OCD can often endure increased disruptions to daily life, including the inability to participate at places of work or in school \cite{matt_wood_medication_2018}.}
\subsection{Current Solutions}
\textnormal{Though OCD has been established as a severely debilitating condition, treatments and knowledge of the disorder are still developing. Current treatments involve utilizing Selective Serotonin Reuptake Inhibitors (SSRIs) as prior research hypothesized serotonin to be a target for effective treatment of the disorder. However, when tested, 40 to 60 percent of patients noticed zero to partial improvements in their symptoms \cite{kellner_drug_2010}. Low success rates with this specific class of drug demonstrate a lack of understanding of targeting the source of OCD; however, the issue is further convoluted by SSRIs remaining the most common choice for medicating OCD patients \cite{xu_optimal_2021}. Rather than focusing on serotonin-based solutions, glutamate-based treatments have risen as a novel approach to understanding the causal factors involved. For instance, these treatments have recently been used to target NMDA receptors for OCD patients \cite{karthik_investigating_2020}. However, findings based on glutamate could be difficult to generalize as the substance is abundant in the brain and underpins various aspects of learning and memory; therefore, unintended consequences could abound as it is not clear that the substance could be targeted specifically for OCD with the exclusion of its other functions. Our work is important in isolating a more narrowed aspect of glutamate for the causation of OCD.}
\subsection{Glutamate}
\textnormal{Glutamate is an excitatory neurotransmitter, with the function of stimulating nerve cells that send a chemical message between different nerve cells. Glutamate is made from glial cells in the brain and is recycled as the older glutamate is simply refreshed with new glutamate naturally. Beyond serving the different trigger actions, glutamate also helps to process gamma-aminobutyric acid, which is another neurotransmitter to calm the brain. In the body, glutamate serves to enhance learning and memory, energy sources for brain cells, chemical messengers, sleep-wake cycles, and pain signaling \cite{cleveland}. Therefore, in the scope of OCD, where obsessive behaviors—such as constant checking—are prevalent, the involvement of glutamate becomes a potential target for therapy. At Ruhr University in Germany, researchers were able to determine that excessive glutamate led to a higher cerebrospinal fluid level in OCD patients compared to non-OCD patients \cite{ocdf2009}. High levels of glutamate were also observed in OCD patients based on a magnetic resonance spectroscopy scan at Wayne State University \cite{ocdf2009}. To confirm correlation, gene expression data in varying regions of the brain can confirm the up-regulation of metabotropic glutamate receptors (GRM), validating the involvement of glutamate.}

\subsection{T1 resting-state MRI}
\textnormal{In the case of GRM, T1 resting-state MRI scans can be utilized for analysis. These scans can identify structural regions of the brain, unnoticeable to the human eye, that deviate across disorders that may correlate with mGluR5. Gene expression analysis based on these scans then provides an outlet to map MRI scans to data points that undergo analysis. However, if no significant features are noted from these MRI scans for OCD, the results may instead indicate support for the \textit{p-factor} theory, which states that all disorders are on a continuum rather than discrete \cite{marshall_hidden_2020}.}
\subsection{Engineering Statement}
\textnormal{Determining the root cause of obsessive-compulsive disorder is highly difficult. At a high level, it is often difficult to differentiate obsessive-compulsive disorder from major depressive disorder and schizophrenia. As a result, the overall aim of this project was to design models for each disorder, develop activation heatmaps, and extract regions of interest. These models serve as a stepping stone in reaching the significance of GRM in OCD patients. To supplement these models, gene expression analysis was conducted afterward to determine the involvement of mGluR5, encoded by GRM, in OCD patients.}
\subsection{Engineering Objective}
\textnormal{OCD diagnosis is currently understudied and misunderstood within the field of neuroscience. Based on these observations, a few main objectives were enacted:
\begin{itemize}
  \item Obj. 1a: Construct individual CNNs with guidance from pre-trained networks for OCD, MDD, and Schizophrenia respectively, with accuracy rates of at least 80\%. 
  \item Obj. 1b: Develop activation heatmaps, demonstrating regions of interest unique to each disorder.
  \item Obj. 1c: Perform gene expression analysis on T1 resting-state MRIs with transcriptomics. 
  \item Obj. 1d: Provide an online web application to allow patients to receive data from the model's details in Obj. 1a, as well as to feed more data into the models.
\end{itemize}
Research on obsessive-compulsive disorder regarding root causes is still misunderstood because of the privatization of datasets and knowledge. As a result, Obj. 1d provides a method to aid future research in being able to expand upon past knowledge in a more accessible and reasonable manner. Obj. 1d will also allow patients to receive helpful metrics free of charge.
}

\section{Methodology}
\subsection{Role of Student vs. Mentor}
\textnormal{Over 6 months, I conducted work within the general area of machine learning. For this project, I take accountability for the work done with modeling and results. I received guidance from my advisors in developing my ideas and how to probe further into findings. I also received assistance in developing a mastery of machine learning-based technologies from my mentors.}
\subsection{Equipment and Materials}
\textnormal{To achieve the objectives outlined, a plethora of resources were utilized. Models were constructed on Python 3.10.0 with TensorFlow Keras. Within these models, numerous technologies were required for development: SimplelTK, Pandas, Matplotlib, NumPy, MedPy, Skimage, seaborn, ResNet50, MobileNet, Scikit-learn, NPM, node.js, AWS, and Imaging-transcriptomics.}
\textnormal{Furthermore, in terms of hardware, this project was conducted using a 2022 Apple MacBook Pro (M2 processor, 8GB ram). Additionally, parts of the models were constructed on a 2022 Apple MacBook Mini (M2 processor, 8GB ram).}
\subsection{Datasets}
\textnormal{To construct models for the disorders in question, T1 resting-state MRI is required in plentiful amounts for each. Data acquired for MDD was sourced from Bezmaternykh D.D et al. 2021. This database contains 72 patients with T1 resting-state MRI scans. The repetition time for these scans was 2.5 seconds with a 90-degree flip angle. T1 resting-state MRI scans for schizophrenia were acquired from Poldrack, R. et al. 2021. The dataset from Poldrack, R. et al. provided MRI scans for several disorders, though only schizophrenia and control patients were used. OCD T1 fractional anisotropy scans were sourced from Kim, Seung-Goo, et al., 2015. Each of these datasets was pre-processed with NumPy resizing techniques to fit the data to size requirements. Afterward, scikit-learn distributed the datasets into the configuration of 70\% train, and 30\% test \cite{nguyen_influence_2021}. Missing anatomy sub-folders were excluded before model creation. Incorrectly sized images or mistimed scans were also subject to removal. AWS and npm were utilized for large file downloads from the databases.}

\begin{Figure}
 \centering
 \includegraphics[width=\linewidth]{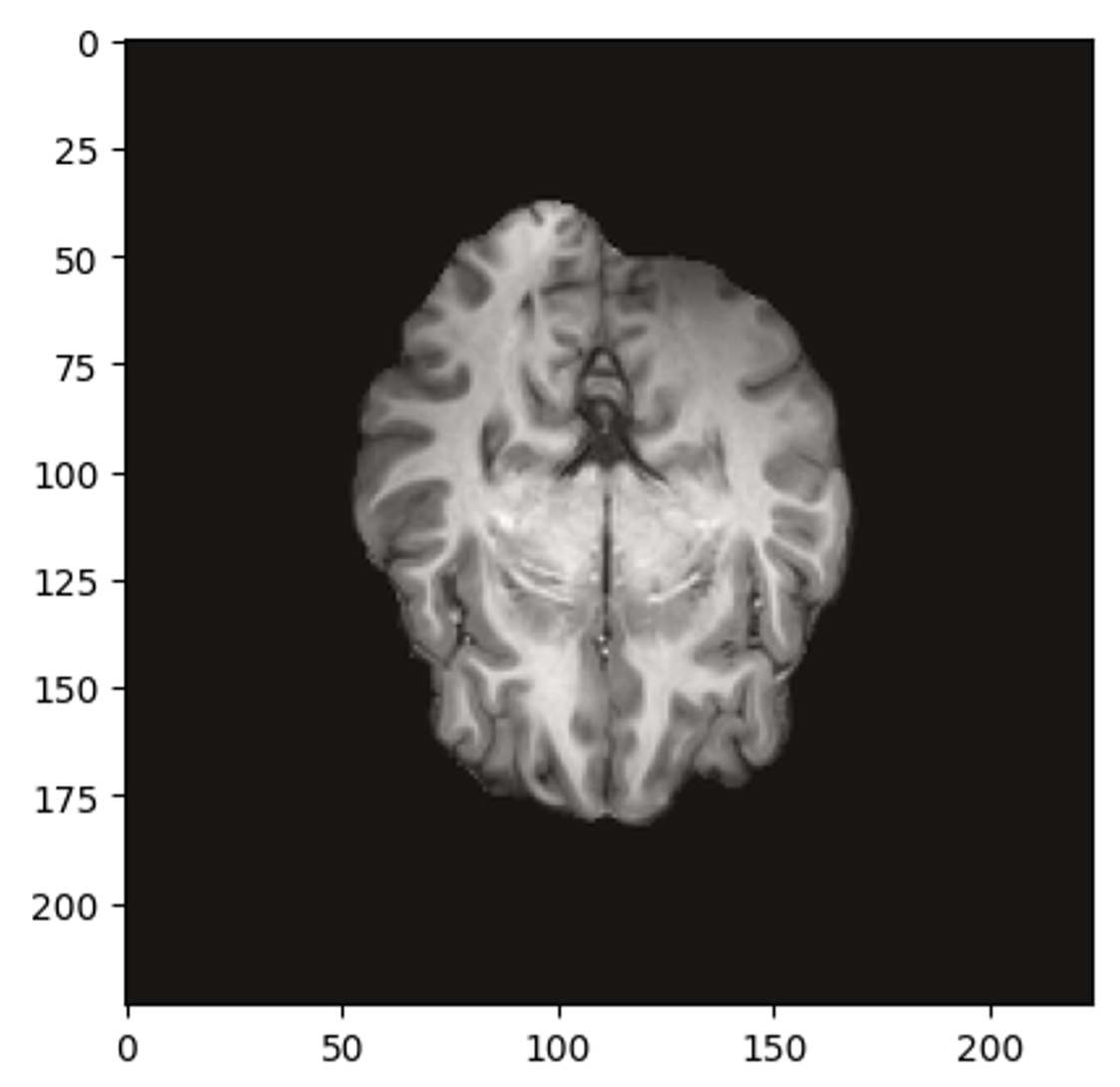}
 \captionof{figure}{TRS-MRI Sample: An example of a 2D resting-state MRI scan acquired from the UCLA Consortium and pre-processed with NumPy.}
\end{Figure}
\subsection{Novel 2D CNNs}
\textnormal{Novel 2D convolutional neural networks were established for each disorder to provide proof of concept for further testing. Each novel CNN was composed of a sequential base, containing pooling, batch, and dropout layers to condense T1 MRI slices into a 1x1 matrix within the sigmoid layer. The model was compiled with an Adam optimizer to reduce computation time \cite{yi_effective_2020}. Furthermore, this optimization allowed for an easier load during this proof of concept. Each model compiled over 25 to 100 epochs with a batch size of 32, amounting to roughly 11.7 hours of runtime per trial.}
\subsection{Optimized Neural Networks}
\textnormal{Pre-trained frameworks were used post-confirmation of functioning novel models. Slices were re-scaled by a scale factor of 0.874 per ResNet50 requirements. During pre-processing, slices with differentiating time stamps due to issues within the scan were excluded. Furthermore, T1 MRI scans noted as corrupted or missing by the primary author of each dataset were also excluded. During the usage of ResNet50, the ram requirement for 23.2 billion parameters surpassed the number of resources available with the given hardware. As a result, the central 40 scans per disorder model were utilized to allow for model compilation. As per ResNet requirements, the coordinate arrays from STIK were stacked by 3 and fitted to the respective shape per disorder (X, Y, 3). The default weight of ImageNet from ResNet50 was in use for modeling. The model ran with a base learning rate of 0.001 throughout 30 epochs. The first 10 epochs were run without interference from ResNet weights whereas pre-trained models and the activation of ImageNet executed from epoch 10 onwards. A similar approach was adopted to make use of MobileNet.}
\subsection{Activation Heatmaps}
\textnormal{Activation heatmaps of the neural network models were created following model development. To construct heatmaps, testing data was classified based on the prediction attribute of the model. Afterward, the layer at index -1 was extracted to obtain weights of size 2048. Based on the usage of ResNet, the final conv layer, conv5block3out, was extracted at sizes 7, 7, and 1. This matrix was then resized to a T1 resting-state size of 189 X 189 X 2048 to provide an overlay of the heatmap onto the original MRI. A Cmap of “jet” was incorporated at an alpha level of 0.5 to highlight regions of importance per each disorder. }
\subsection{Statistical Tests}
\textnormal{To analyze the performance of the models, the F1 score, and confusion matrix were generated. Additionally, the Matthews Correlation Coefficient score was analyzed. These metrics were selected based on their specificity to binary classification models \cite{hutchison_probabilistic_2005}. Classifying the results of a model using traditional statistics was averted because of variations in data sets and methodology. The F1 score was calculated as follows:
}
\[
F1 = \frac{2 \cdot \text{precision} \cdot \text{recall}}{\text{precision} + \text{recall}}
\]
\section{Results}
\subsection{Dataset Creation}
\textnormal{In the initial stages of this project, large datasets from various sources were compiled before analysis. In total, over 150 sub-sessions of T1 resting-state scans were developed through the course of this project. In light of limited data accessibility, acquiring data of this amount presents a means for future use in finding MRI-related datasets.}
\subsection{Novel 2D CNNs Predictions}
\textnormal{Through the course of model development, the novel 2D CNN was utilized as a means of providing reliable evidence of functionality before moving forward. In the case of MDD, the model accuracy approached 99.99\% as, the validation accuracy fluctuated through 100 epochs as demonstrated by Figure 2. These results are then shown by a t-SNE plot in Figure 3 —a visualization technique that demonstrates the low-dimensional representation of high-dimensional of predictions. These plots will reveal the underlying patterns, clusters, or similarities among the MRI scans. Furthermore, the statistical analysis technique of a confusion matrix was utilized for the novel models as shown in Figure 4. This measure allows the generalization of the overall effectiveness of the model at a glance. Precision, recall, F1, and detailed analysis were constructed for definite pre-trained models. These results enabled the usage of pre-trained neural networks. Furthermore, the novel 2D CNN constructed for schizophrenia also yielded promising results with a model accuracy of 99.77\% and a validation accuracy of 82.08\%. The final accuracy fell to around 79\% but the model restores best weights on run. Figure 4 demonstrates the confusion matrix for schizophrenia. Based on the results provided by ResNet50, a novel model was not constructed for OCD.}

\begin{Figure}
 \centering
 \includegraphics[scale=0.35]{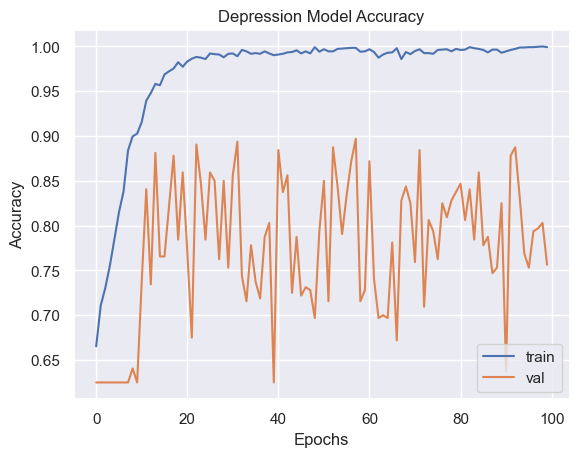}
 \captionof{figure}{MDD Validation: Validation and model accuracy metrics over the interval of 100 epochs for MDD.}
\end{Figure}
\begin{Figure}
 \centering
 \includegraphics[scale=0.22]{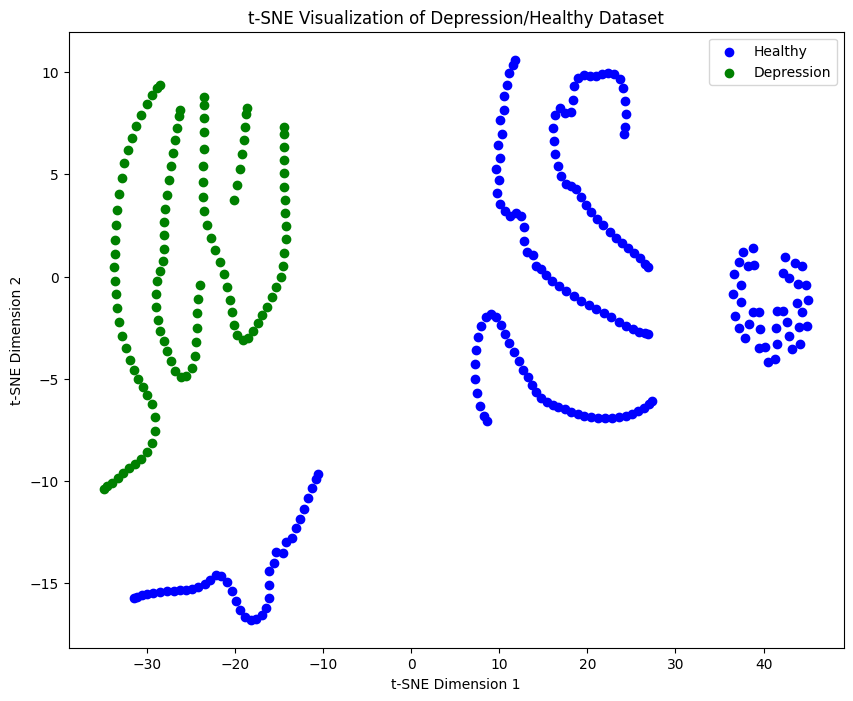}
 \captionof{figure}{t-SNE Visualization: A low-dimensional representation of high-dimension, complex MDD MRI slices.}
\end{Figure}
\begin{Figure}
 \centering
 \includegraphics[scale=0.25]{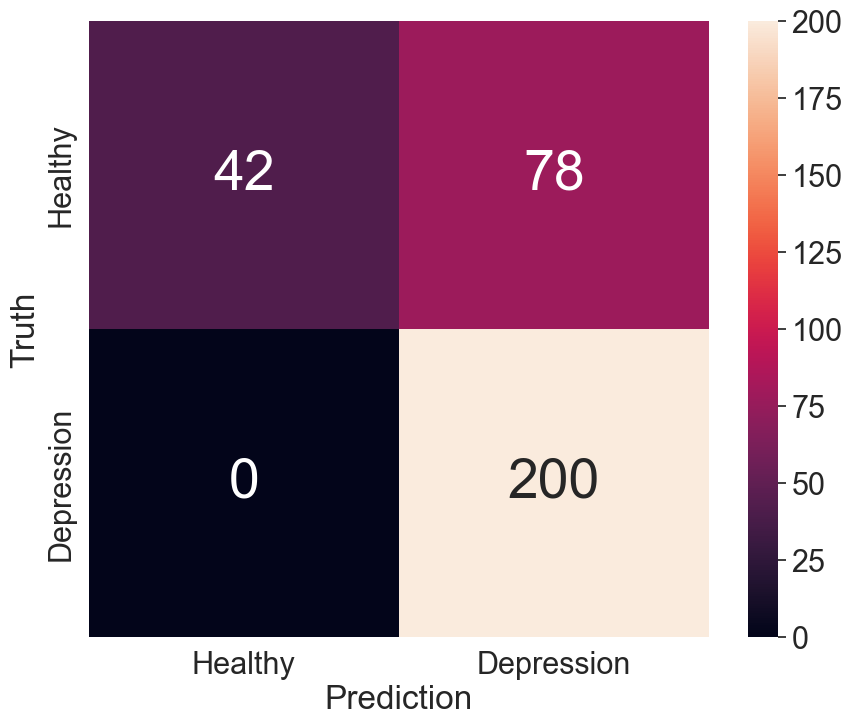}
 \captionof{figure}{MDD Confusion Matrix: Confusion matrix statistical analysis for MDD model performance and accuracy.}
\end{Figure}
\newpage
\begin{Figure}
 \centering
 \includegraphics[scale=0.25]{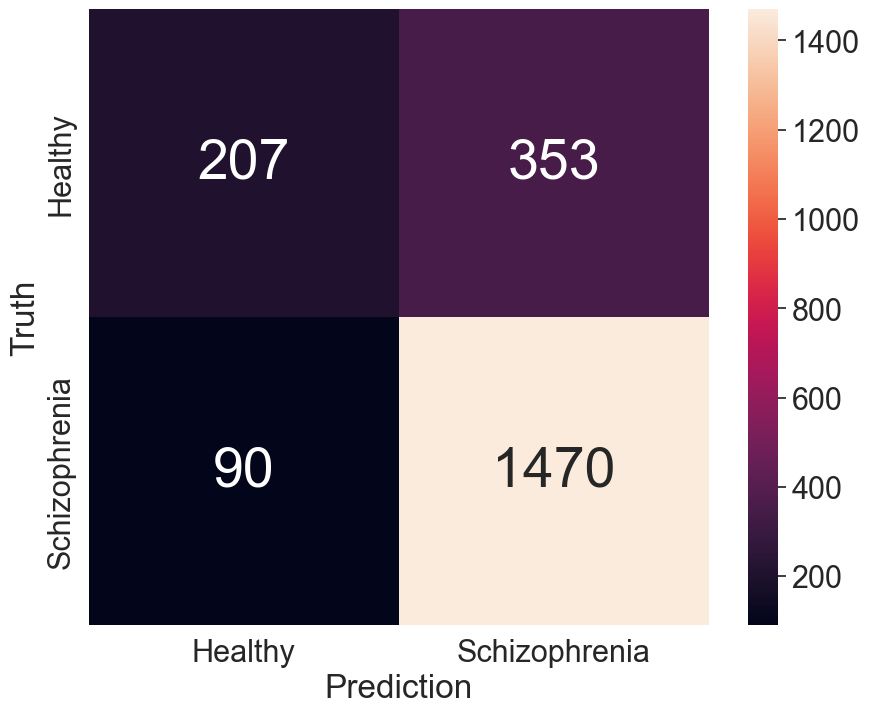}
 \captionof{figure}{SZD Confusion Matrix: Statistical analysis chart based on schizophrenia model performance and accuracy.}

\end{Figure}
\subsection{Optimized Neural Networks Predictions}
\textnormal{After the creation of novel networks, ResNet50 and MobileNet were investigated to achieve higher rates of validation accuracy. To visualize the variance created by the usage of a pre-trained network, a green fine-tuning line is incorporated at epoch 10. For MDD, training approached the limit of 1.00 while the validation accuracy stabilized at 88.75\%. Cross entropy readings fluctuated but reached a natural deviation after epoch 15. Further insight into the overall performance of the model is provided by the calculated statistical scores in Table 1. These scores were all produced for the pre-trained models to protect against potential overfitting in the novel 2D CNNs. }

\begin{center}
\captionof{table}{Statistical Measures Based on RESNET{\_} mdd{\_} resting{\_} state{\_} dataset{\_} t1{\_} 2d.h5}
\vspace{10pt}
\begin{tabular}{c|c}
\hline
Metric & Value \\
\hline
\rowcolor{Gray}
Precision & 0.6000 \\
F1 Score & 0.7423 \\
\rowcolor{Gray}
Mathews Correlation Coefficient & 0.6774 \\
Sensitivity & 0.9730 \\
\rowcolor{Gray}
Specificity & 0.8049
\end{tabular}
\end{center}

\begin{Figure}
 \centering
 \includegraphics[scale=0.35]{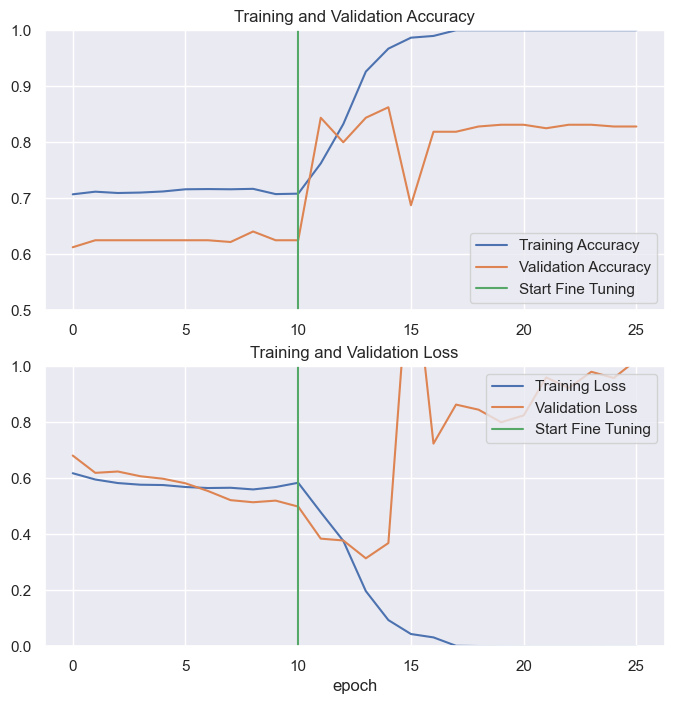}
 \captionof{figure}{ResNet50 MDD Validation: ResNet50 validation and cross entropy with respect to change in epoch.
}
\end{Figure}
\begin{Figure}
 \centering
 \includegraphics[scale=0.25]{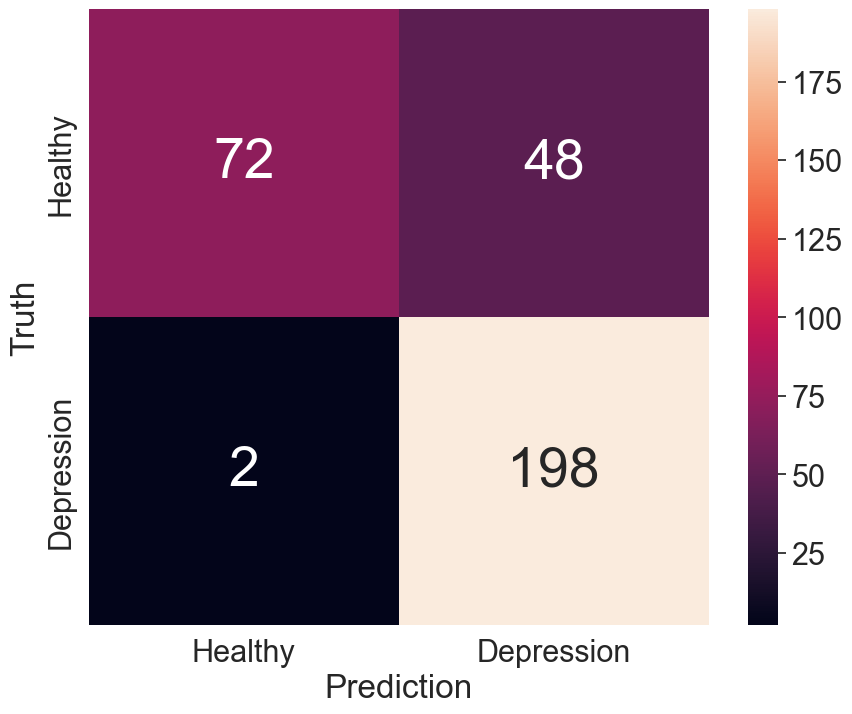}
 \captionof{figure}{ResNet50 MDD Confusion Matrix:
ResNet50 confusion matrix based on model performance.
}

\end{Figure}

\textnormal{Figure 6 demonstrates the change in accuracy and cross-entropy over time, while Figure 7 is the corresponding matrix results. Similar results were produced regarding the Schizophrenia ResNet50 model. Patient 126 was excluded from the sample due to difference in numpy dimensions. Afterward, the remaining scans were cross-validated for shape, as shown in Figure 10. The model yielded a validation accuracy of 80.5\% and a model accuracy that approached 99.9\%. Before fine-tuning, the model averages a validation accuracy of around 70\%. After weight and pre-train activation, the model climbs in terms of model accuracy, though the validation accuracy remains stable. The final classifications of patients can be seen in the confusion matrix of Figure 9. The t-SNE plot in Figure 8 demonstrates the model performance with mixed clusters. The resulting statistical measures are shown in Table 2.}

\begin{center}
\captionof{table}{Statistical Measures Based on RESNET{\_} szd{\_} resting{\_} state{\_} dataset{\_} t1{\_} 2d.h5}
\vspace{10pt}
\begin{tabular}{c|c}
\hline
Metric & Value \\
\hline
\rowcolor{Gray}
Precision & 0.5786 \\
F1 Score & 0.6166 \\
\rowcolor{Gray}
Mathews Correlation Coefficient & 0.4928 \\
Sensitivity & 0.6599 \\
\rowcolor{Gray}
Specificity & 0.8551
\end{tabular}
\end{center}

\begin{Figure}
 \centering
 \includegraphics[scale=0.3]{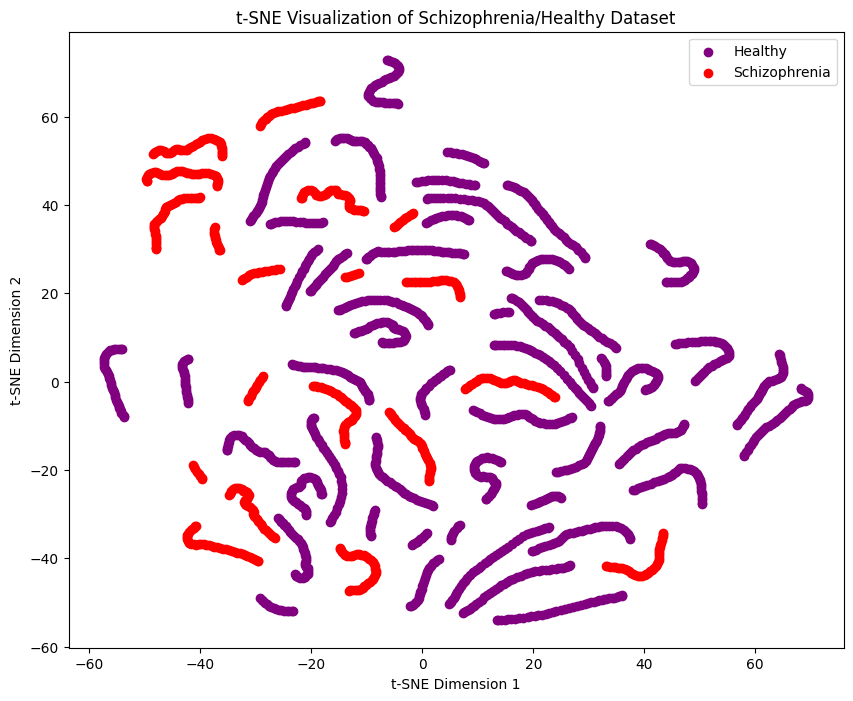}
 \captionof{figure}{SZD ResNet50 t-SNE Validation: A low-dimensional representation of high-dimension, complex SZD MRI slices.}
\end{Figure}
\begin{Figure}
 \centering
 \includegraphics[scale=0.25]{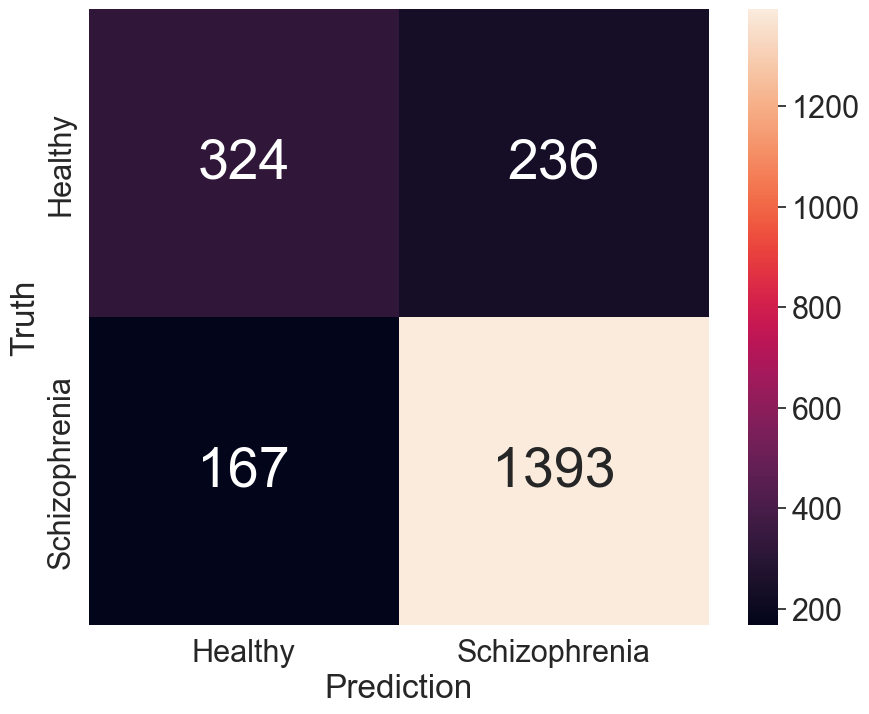}
 \captionof{figure}{ResNet50 SZD Confusion Matrix:
Confusion matrix generation for performance of the ResNet50 schizophrenia model.}
\vspace{-1.0em}
\end{Figure}
\begin{Figure}
 \centering
 \includegraphics[width=\linewidth]{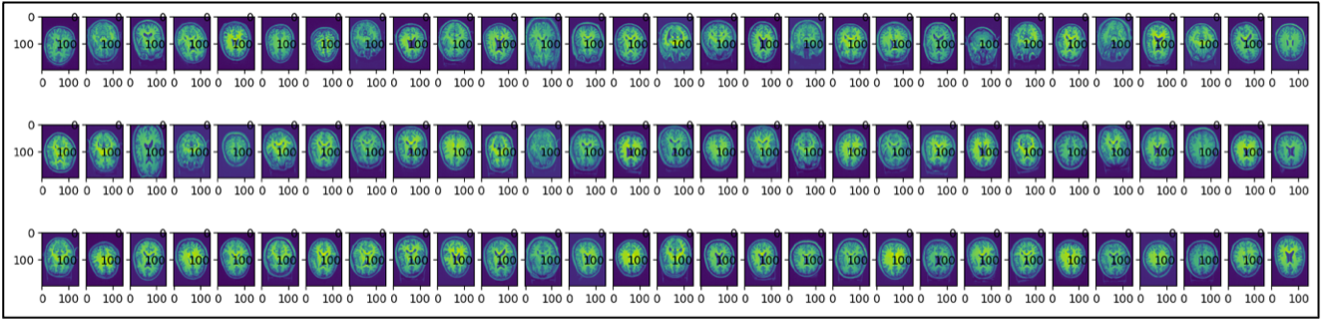}
 \captionof{figure}{Verifying SZN TRS-MRI Data:
Remaining T1 resting-state MRI scans after removal of sub-126. Shape validation was conducted on these scans for dimensions (151, 40, 199).
}
\end{Figure}
\textnormal{When conducting a model for OCD, both the model and validation accuracies yielded levels around around 54.44\%. Furthermore, the confusion matrix resulted in 0 predicted slices of OCD while the remaining 120 were classified as healthy by the model.}

\begin{Figure}
 \centering
 \includegraphics[scale=0.2]{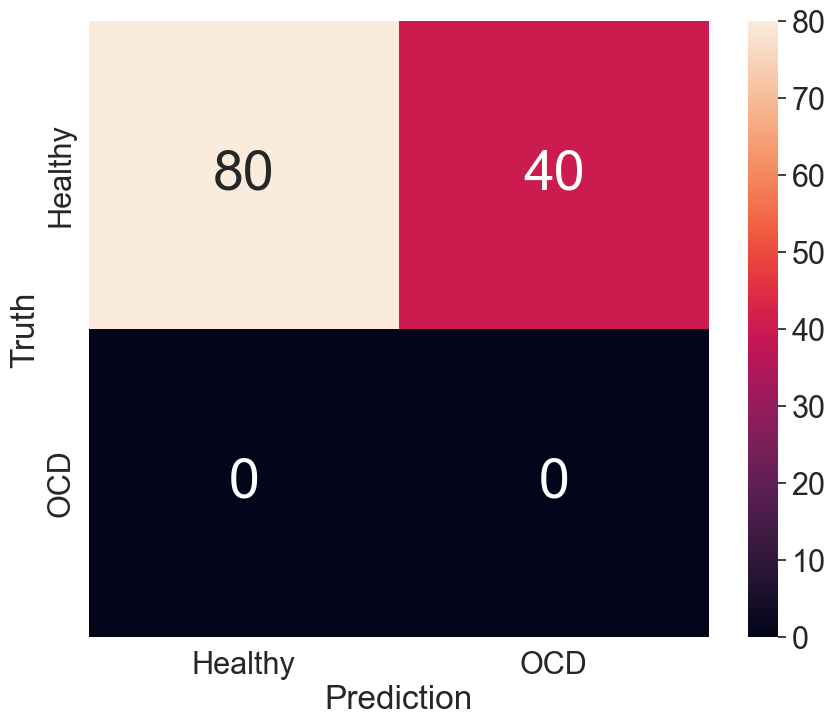}\hfill
 \includegraphics[scale=0.43]{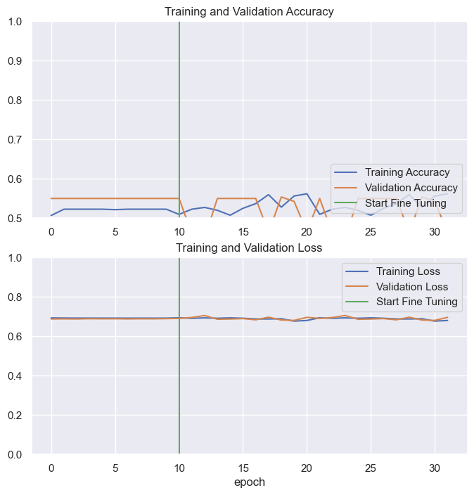}
 \captionof{figure}{(left) ResNet50 validation and cross entropy with respect to change in epoch for OCD (right) ResNet50 validation and cross entropy with respect to change in epoch for OCD.
}
\end{Figure}

\subsection{Activation Heatmap Findings}
\textnormal{After confirming relatively fair accuracy within the models constructed, activation heatmaps were constructed with both patients diagnosed with a disease and healthy patients. Scaled convolution layers were placed over the original image to extract the regions of interest. Cmap “jet” outputted the heatmaps as zones. Red and orange zones are representative of areas in which the neural network detected differences in structural patterns. Heatmaps were not outputted for OCD. Figure 12 depicts the heatmap configuration for MDD while Figure 13 demonstrates the configuration for schizophrenia. Partial scans from MDD and SZD are displayed; all heatmaps can be found in appendices B and C.}

\begin{Figure}
 \centering
 \includegraphics[scale=0.6]{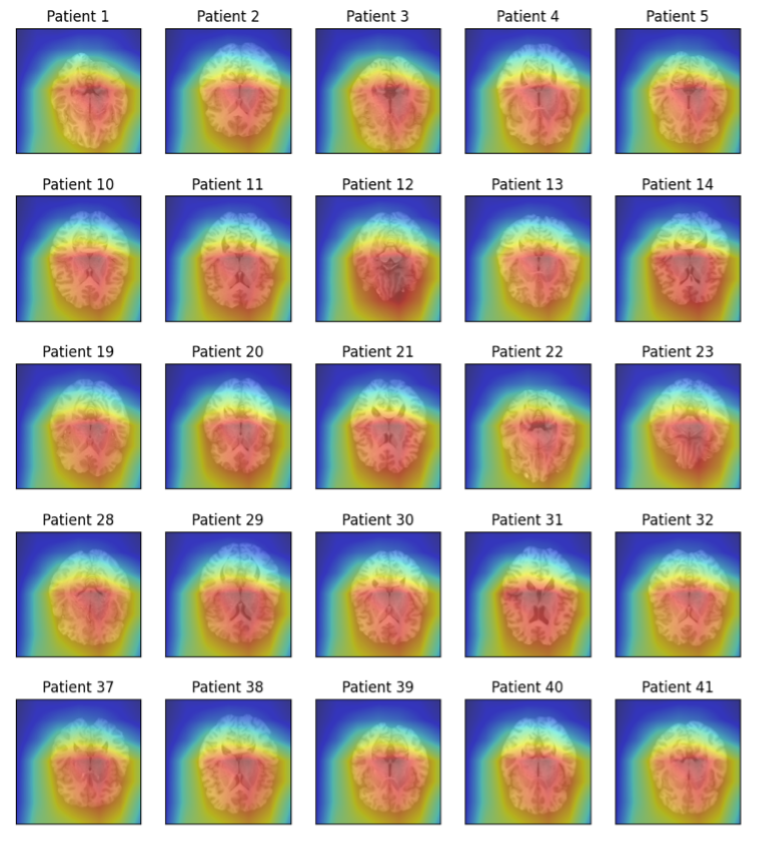}\hfill
 \captionof{figure}{MDD Activation Heatmaps:
The figures above demonstrate the activation heatmaps cast onto the 72 patients within the MDD model.}
\vspace{-1.0em}
\end{Figure}
\begin{Figure}
 \centering
 \includegraphics[scale=0.6]{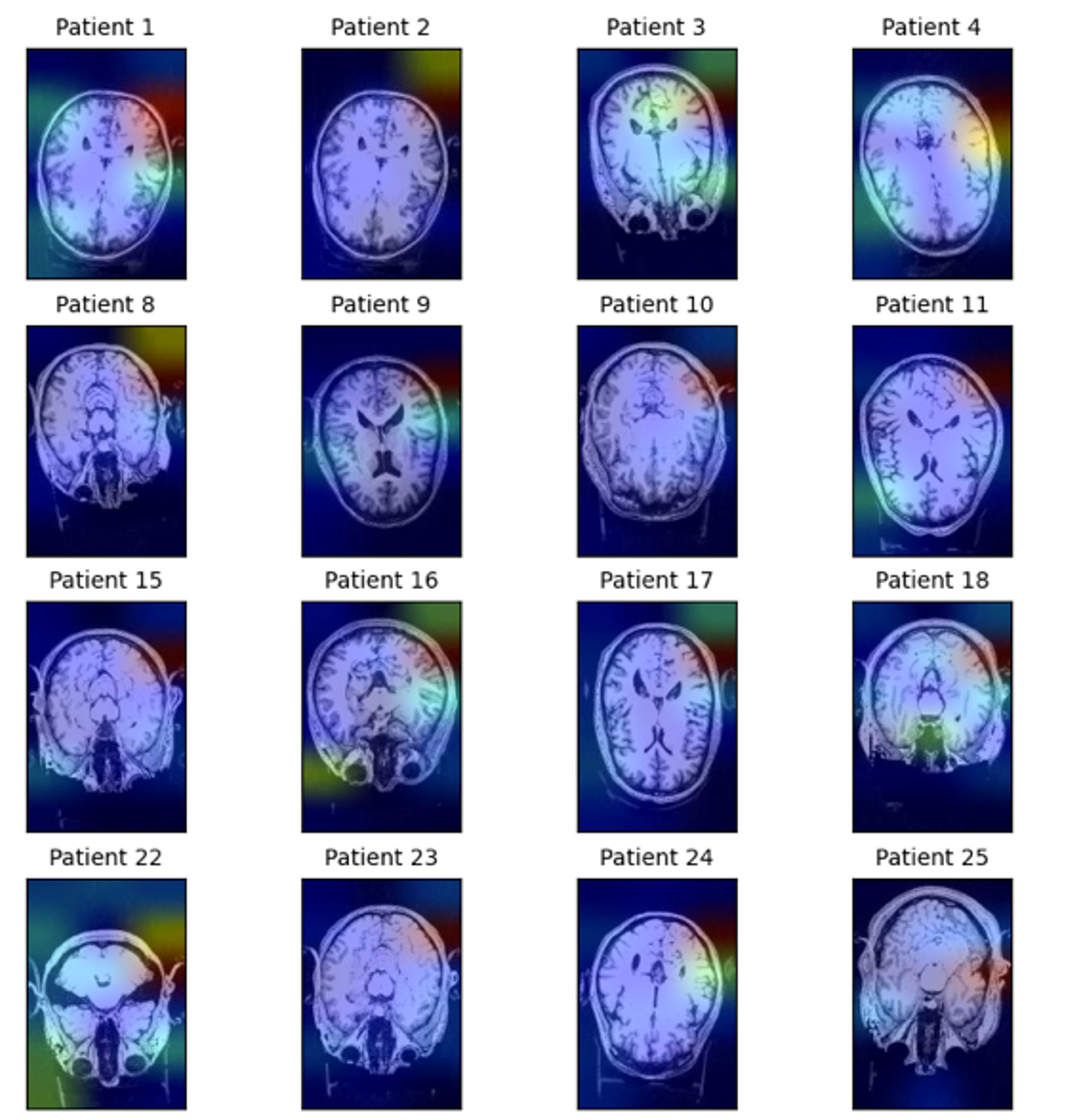}\hfill
 \captionof{figure}{Schizophrenia Activation Heatmaps:
The figures above demonstrate the activation heatmaps for the schizophrenia model. Regions of red are areas of interest in diagnosis.}
\end{Figure}

\section{Discussion}
\textnormal{Determining the root cause of obsessive-compulsive disorder is highly difficult. As a result, the overall aim of this project was to design models for each disorder, develop activation heatmaps, and extract regions of interest. From these models, heatmaps became an additional objective to provide an overview of regions of interest for the 2D CNN. Furthermore, these models were intended to be used in gene expression analysis to identify significant gene encodings that demonstrate higher levels of influence in OCD as compared to other disorders A two-proportion z-test was utilized throughout all model cases due to the nature of the data: varying percentages requiring a metric to be standardized and compared. }
\newline
\newline
\textnormal{Through conducting this study, respective models for MDD, SZD, and OCD have been successful in construction. Each demonstrates an important and different aspect in terms of OCD. The MDD model provided a validation accuracy of 88.75\%. Within the field of computational neuroscience, this accuracy serves to provide more reliable results as compared to other models. When compared to other models, our MDD classification network produced a P-Value of 0.04363, significant at an alpha level of 0.05. Achieving this level of significance allows for the model to be presented as a viable classification method in the future. The t-SNE plot from Figure 3 also depicts separation in clusters of healthy vs. MDD patients. Even within the healthy MRI scans, there are some distinct clusters, which may be an area of future research. However, the primary clusters show that there is a clear factor the model is able to identify when classifying patients. This factor is found when analyzing the heatmaps which demonstrated implications in the corpus callosum. This finding is consistent with past literature where implications in the corpus callosum were found \cite{lee_shape_2020}. By demonstrating further support for the involvement of the corpus callosum, future research can have greater assurance of results based on our work.}
\newline
\newline
\textnormal{The schizophrenia model provided similarly assuring results as with MDD. To test the concept of computational modeling, a novel network was built as a primary means. Though we theorized that shifting to ResNet50 to bolster accuracy with ImageNet, the novel network provided a greater level of accuracy at 82.08\% in comparison to the optimized network. This novel network was not optimized per layer, meaning that future work has the potential to bolster accuracy rates with a different framework for transfer learning. ResNet50's \textit{include-top} parameter was declared as \textit{false} meaning that it effectively acted as a transfer learning model; however, in the future VGG16 may be explored as a different means  of transfer learning. The t-SNE plot from Figure 8 demonstrates a high level of overlap between the healthy and SZD patients. This overlap may suggest why the model validation accuracy stalled at around 80\%. At this range, due to overlapping clusters of features, improvement may hae been different. Though for the separations identified, the GradCam heatmaps demonstrate implications within the right frontal lobe. Determining the importance of the right frontal lobe, or more broadly the frontal lobe, aligns with previous works within the field of neuroscience that also find particular importance in the right frontal lobe \cite{a_frontal_2016}. These results can narrow down future research regarding schizophrenia—a contribution that resides outside of the topic of focus.}
\newline
\newline
\textnormal{Unlike MDD and SZD, the OCD novel and optimized networks failed to predict healthy patients from OCD patients. When first observing the results, this failure appeared to be the cause of TRS-MRI distortions during the scans from the dataset; however, further analysis demonstrated that the framework was able to sufficiently read the coordinate points of scans. Furthermore, the authors of the data with FSL \cite{kim_alterations_2015}. As a result, OCD is evidenced to possess no unique characteristics that define it differently from a healthy brain, leading to the inability to predict a diagnosis. As demonstrated by the confusion matrix in Figure 11 (left), the model outputted that all scan slices were healthy patients. If no defining characteristic can be found by the classification network, it must default to healthy as per the logistics of binary classification. These findings are consistent with a meta-analysis conducted by the ADAA \cite{amy_rapp_how_nodate}.  The authors found that across 100 studies collected, they were unable to find a consistency of OCD function implications that could be a cause of a cognitive aspect. In addition, they note that due to these inconsistencies in findings across studies, pushing toward adequate treatments in the future becomes increasingly difficult. Our model summarizes these descriptions by demonstrating that for multiple reasons, the model fails to provide a viable method to predict future cases due to comorbidity and a lack of differentiability in scans. Therefore, the objective for OCD TRS-MRI slices was not satisfied, but a deeper understanding of the field and its reflection within the model was understood.}
\newline
\newline
\textnormal{Though SZD and MDD neural networks point to specific regions of scan slices, the overall findings of this project further the \textit{p-factor theory}. This theory alludes to the fact that all psychopathological disorders can be generalized under one umbrella. Authors of this paper further contribute to the concept of being inability to utilize TRS-MRIs instating that groups attempting to neglect the \textit{p-factor} and find regions of interest tend to create contradictory results to preexisting studies \cite{marshall_hidden_2020}. Understanding disorders with lacking knowledge will require utilization of the \textit{p-factor}. OCD lacks fundamental understanding; though, the use of the \textit{p-factor} presents a viable solution to treat OCD as well as other under-studied disorders in the future.}
\vspace{-0.8em}
\subsection{Limitations}
\textnormal{Throughout conducting this project, the major limitation faced was computing power. Due to the 8GB RAM limit which resulted in computer crashes, only certain subsets of data were able to be processed. Surpassing this barrier will allow for the usage of deep transfer learning in the form of a support vector machine in the future. Data acquisition arose as another limitation and restricted the project to publicly available datasets. For instance, this study intended to utilize PET scan data in the initial stages but switched to TRS-MRI due to data availability.}
\vspace{-0.8em}
\subsection{Future Research}
\textnormal{Future research may involve the optimization of the models constructed in this paper. Though the accuracy rates were promising, model validation can be presented with deep learning methods. Furthermore, in terms of SZD and MDD, rather than being stratified into two separate disorders, future work may focus on condensing these models under one network that can differentiate the disorders across multiple classes. The creation of this combined model has the potential to provide more supplementary evidence to the \textit{p-factor.} Providing further evidence for the \textit{p-factor} is essential to develop more effective and widely used treatments in the future.}
\section{Conclusion}
\textnormal{Our engineering project served to identify the differentiation factor between OCD and disorders comorbid with OCD. As a result, SZD and MDD were selected for cross-comparative analysis. From there, the project served to generate heatmaps as well as a gene expression analysis model to determine the significance of gene encodings across the disorders. These objectives were set on the basis that mGluR5 was implicated in those suffering from OCD. To conduct the study, this study we focused on using convolutional neural networks to classify disorders with the means of novel, ResNet5, and MobileNet models. Afterward, heatmaps were generated following ResNet50 guidelines and with GradCam. However, before approaching the gene expression analysis stage, the TRS-MRI scans were unable to predict cases of OCD based on slices. As a result, we interpreted the model as unable to predict OCD cases because of multiple regions were connected to OCD as a neurological disorder. Though diverting from the original objective of finding the differentiation factor for OCD, the lack of prediction still points to a fundamental piece of motivation for this project—a lacking form of treatment for OCD. Lacking predictions, though, still provided useful insight as they pointed towards a theory known as the \textit{p- factor} \cite{marshall_hidden_2020}. This theory states that within the field of neuroscience, researchers are unable to find distinctive characteristics respective to individual disorders because these disorders are not discrete and share many overlapping regions of implication. Rather, they  Therefore, these disorders fall onto a continuum that must be treated in an empirical manner rather than case-by-case. As a result, it becomes clear that searching for universal characteristics that can facilitate diagnosis is paramount. Though a singular factor was unable to be uncovered through this project, further support for the \textit{p-factor} was found, providing further evidence that disorders are on a continuum rather than discrete. These findings guide future research and provide a feasible theory and mode for finding a treatment for OCD.}
\section*{Acknowledgments}
I would like to give special thanks to Dr. Kevin Crowthers and Mr. Nicholas Medeiros for their support throughout the journey of this research project. Additionally, I am grateful for the help provided by Varshini Ramanathan of Tufts University as well as Revathi Ravi of Massachusetts General Hospital for their guidance in topic-specific knowledge as well as with data needs. Finally, I am thankful for the guidance provided by Joseph Yu in aiding my understanding of machine learning concepts for this project.
\bibliographystyle{unsrt}  
\bibliography{paper}  
\section{Code Availability}
\textnormal{The code written and used in the analysis for this project can be found \href{https://github.com/Tarune28/Modeling-T1-Resting-State-MRI-Variants-Using-Convolutional-Neural-Networks-in-Diagnosis-of-OCD}{here }(clickable). Code used may be reused and modified from this project with appropriate attribution. Each subfolder identifies the disorder by the name used in this study. The links section of the repository contains a re-direct to an application where models can be trialed live. Further information can be found in the README.md file.}
\section{Model Availability}
\textnormal{All of the models referred to in this paper can be found \href{https://gitlab03.wpi.edu/teswar/trs-mri-models}{here} (clickable). Folders are labeled by disorder and model. The corresponding accuracy for each model can be found in Model-Info.tsv file. Each of the models presented in this paper has multiple versions and all versions are stored at this repository.}
\section{Appendices}
\subsection{Appendix A: Limitations and Assumptions}
\underline{Limitations:}
\begin{itemize}
\item Utilization of more memory-requiring models such as SVM was limited. 
\item Training based on fractional anisotropy scan data rather than TRS-MRI in relation to the OCD model.
\item Computing power was limited to 8GB RAM.
\item Limitation to TRS-MRI scans due to motion distortion in fMRI scan data.
\end{itemize}

\underline{Assumptions:}
\begin{itemize}
\item Analysis using FSL conducted by Kim, Seung-Goo, et al., 2015 was accurately conducted for Eddy correction.
\item Variations in heatmap outputs are solely due to the use of the global pooling layer.
\item Central 40 TRS-MRI scans pertain the most relevance to neural network models
\end{itemize}

\subsection{Appendix B: All MDD Heatmaps}
\textnormal{The following images show all of the MDD heatmaps produced with a Cmap overlaid to demonstrate the regions of interest highlighted in this paper.}
\begin{Figure}
\centering
 \includegraphics[scale=0.3]{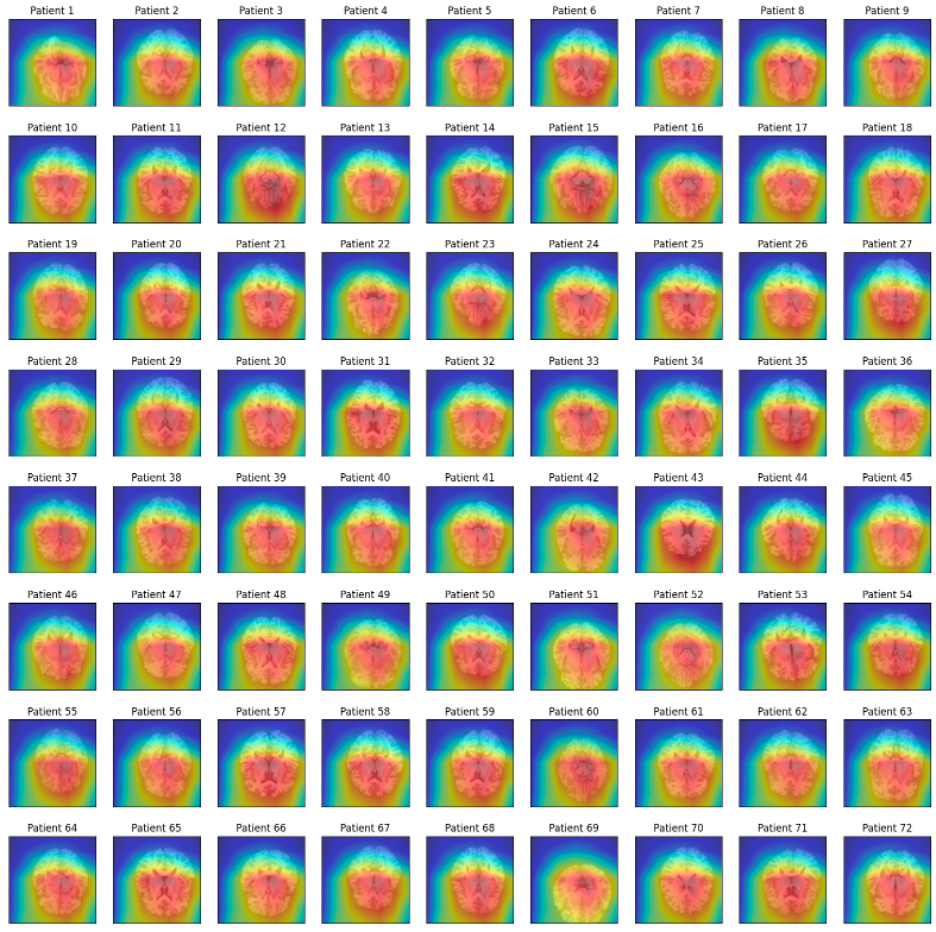}
\end{Figure}
\subsection{Appendix C: All SZD Heatmaps}
\textnormal{In the T1 resting-state MRI scans below, the GradCam results of the model for SZD are shown.}
\begin{Figure}
\centering
 \includegraphics[scale=0.35]{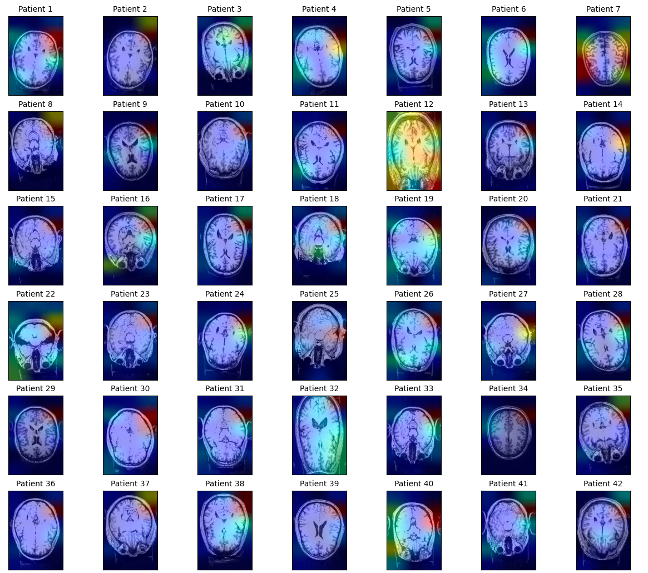}
 \includegraphics[scale=0.35]{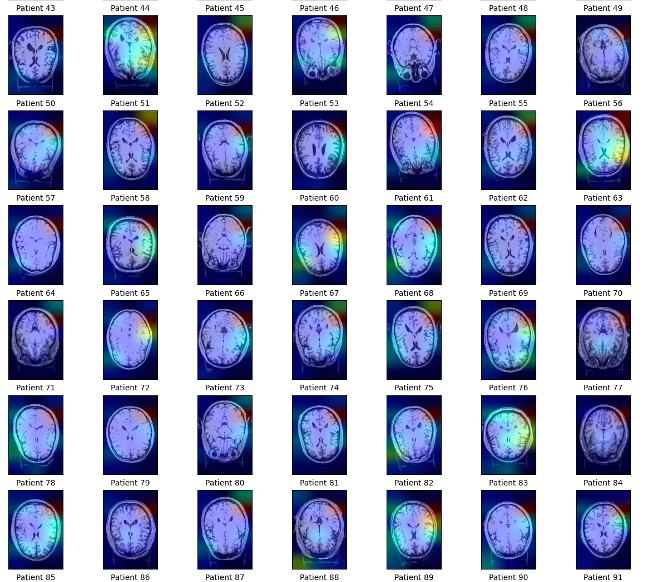}
\end{Figure}
\begin{Figure}
\centering
 
 \includegraphics[scale=0.35]{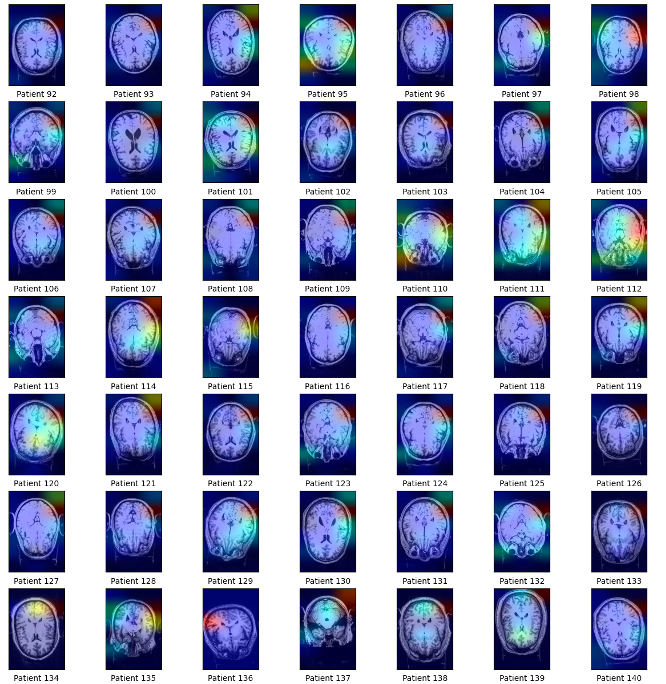}
\end{Figure}
\begin{Figure}
\centering 
 \includegraphics[scale=0.35]{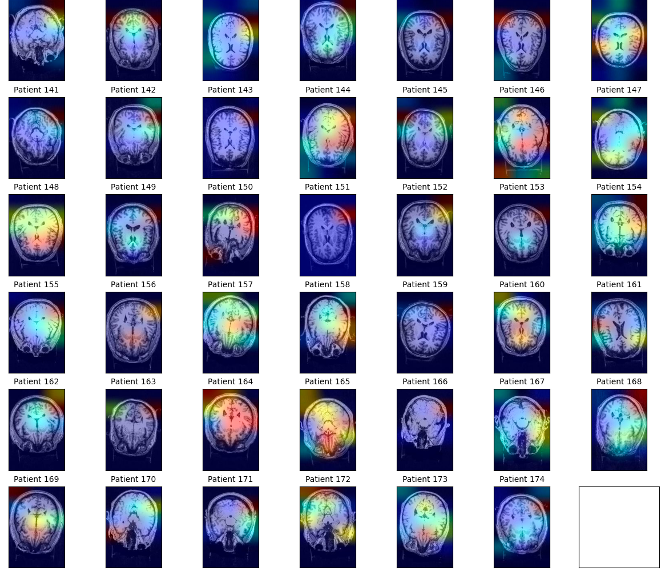}
\end{Figure}
\newpage

\end{multicols}
\end{document}